\documentstyle[epsfig]{mn}
\newif\ifAMStwofonts

\ifoldfss
  \newcommand{\rmn}[1] {{\rm #1}}

  \ifCUPmtlplainloaded \else
    \NewTextAlphabet{textbfit} {cmbxti10} {}
    \NewTextAlphabet{textbfss} {cmssbx10} {}
    \NewMathAlphabet{mathbfit} {cmbxti10} {} 
    \NewMathAlphabet{mathbfss} {cmssbx10} {} 
  \fi
  \ifAMStwofonts
    \ifCUPmtlplainloaded \else
      \NewSymbolFont{upmath} {eurm10}
      \NewSymbolFont{AMSa} {msam10}
      \NewMathSymbol{\upi}     {0}{upmath}{19}
      \NewMathSymbol{\umu}     {0}{upmath}{16}
      \NewMathSymbol{\upartial}{0}{upmath}{40}
      \NewMathSymbol{\leqslant}{3}{AMSa}{36}
      \NewMathSymbol{\geqslant}{3}{AMSa}{3E}

      \let\leq=\leqslant 
       
    \fi
  \fi
\fi 

\ifnfssone
  \newmathalphabet{\mathit}
  \addtoversion{normal}{\mathit}{cmr}{m}{it}
  \addtoversion{bold}{\mathit}{cmr}{bx}{it}
  \newcommand{\rmn}[1] {\mathrm{#1}}

  \newmathalphabet{\mathbfit} 
  \addtoversion{normal}{\mathbfit}{cmr}{bx}{it}
  \addtoversion{bold}{\mathbfit}{cmr}{bx}{it}
  \newmathalphabet{\mathbfss} 
  \addtoversion{normal}{\mathbfss}{cmss}{bx}{n}
  \addtoversion{bold}{\mathbfss}{cmss}{bx}{n}
  \ifAMStwofonts
    \ifCUPmtlplainloaded \else
      \UseAMStwoboldmath
      \makeatletter
      \new@mathgroup\upmath@group
      \define@mathgroup\mv@normal\upmath@group{eur}{m}{n}
      \define@mathgroup\mv@bold\upmath@group{eur}{b}{n}
      \edef\UPM{\hexnumber\upmath@group}
      \new@mathgroup\amsa@group
      \define@mathgroup\mv@normal\amsa@group{msa}{m}{n}
      \define@mathgroup\mv@bold\amsa@group{msa}{m}{n}
      \edef\AMSa{\hexnumber\amsa@group}
      \makeatother
      \mathchardef\upi="0\UPM19
      \mathchardef\umu="0\UPM16
      \mathchardef\upartial="0\UPM40
      \mathchardef\leqslant="3\AMSa36
      \mathchardef\geqslant="3\AMSa3E

      \let\leq=\leqslant 

    \fi
  \fi
\fi 

\ifnfsstwo
  \newcommand{\rmn}[1] {\mathrm{#1}}

  \DeclareMathAlphabet{\mathbfit}{OT1}{cmr}{bx}{it}
  \SetMathAlphabet\mathbfit{bold}{OT1}{cmr}{bx}{it}
  \DeclareMathAlphabet{\mathbfss}{OT1}{cmss}{bx}{n}
  \SetMathAlphabet\mathbfss{bold}{OT1}{cmss}{bx}{n}
  \ifAMStwofonts
    \ifCUPmtlplainloaded \else
      \DeclareSymbolFont{UPM}{U}{eur}{m}{n}
      \SetSymbolFont{UPM}{bold}{U}{eur}{b}{n}
      \DeclareSymbolFont{AMSa}{U}{msa}{m}{n}
      \DeclareMathSymbol{\upi}{0}{UPM}{"19}
      \DeclareMathSymbol{\umu}{0}{UPM}{"16}
      \DeclareMathSymbol{\upartial}{0}{UPM}{"40}
      \DeclareMathSymbol{\leqslant}{3}{AMSa}{"36}
      \DeclareMathSymbol{\geqslant}{3}{AMSa}{"3E}

      \let\leq=\leqslant 

    \fi
  \fi
\fi 

\ifCUPmtlplainloaded \else
  \ifAMStwofonts \else 
    \def\upi{\pi}
    \def\umu{\mu}
    \def\upartial{\partial}
  \fi
\fi

\date{}

\pagerange{\pageref{firstpage}--\pageref{lastpage}}
\pubyear{2007}

\title[Polaris~B]
{Polaris~B, an optical companion of Polaris ($\alpha $\,UMi) system:
atmospheric parameters, chemical composition, distance and mass}

\author[Usenko \& Klochkova]
{Igor Usenko$^1$, Valentina Klochkova$^2$  \\
$^1$ Astronomical Observatory of Odessa State University, Odessa
           65014, Ukraine\\
$^2$ Special Astrophysical Observatory RAS,
           Nizhnij Arkhyz, Karachaevo-Cherkessia, 369167, Russia\\
E-mail: igus@deneb.odessa.ua\\
E-mail: valenta@sao.ru\\
}

\begin{document}

\onecolumn

\maketitle


\begin{abstract}
We present an analysis of high-resolution spectroscopic observations of
Polaris~B, the optical companion of the Polaris~Ab system. The star has
a radial velocity $V_r$ of  $-16.6$\,km\,s$^{-1}$ to $-18.9$\,km\,s$^{-1}$, and
a projected rotational velocity $v \sin i$\,=\,110 km\,s$^{-1}$. The
derived atmosphere parameters are: $T_{\rmn{eff}}$=6900 K;
$\log g$\,=\,4.3; $V_{\rmn{t}}$=2.5\,km\,s$^{-1}$. Polaris~B has elemental
abundances generally similar to those of the Cepheid Polaris~A (Usenko et
al. 2005a), although carbon, sodium and magnesium are close to the solar
values. At a spectral type of $F3\,{\sc v}$ Polaris~B has a luminosity of
3.868 L$_{\sun}$, an absolute magnitude of +3.30$^m$, and a distance of
109.5\,pc. The mass of the star is estimated to be 1.39\,${\mathcal M}_{\sun}$, close
to a mass of 1.38$\pm$0.61\,${\mathcal M}_{\sun}$ for the recently-resolved orbital
periods companion Polaris~Ab observed by Evans et al. (2007).
\end{abstract}


\section{Introduction}

Polaris~B (BD+88$^{\circ}$7; SAO 305) is the nearest optical companion to
the Cepheid binary Polaris~Ab ($\alpha$\,UMi) system. Its visual magnitude
of 8$\lefteqn{.}^m60$, angular distance of 18$^{''}$ from Polaris, and
spectral type of $F3\,{\sc v}$ suggest that it is a physical main-sequence
companion of $\alpha$ UMi (Fernie 1966; Turner 2005). Polaris~A is itself
a small-amplitude Cepheid with a spectroscopic binary companion,
Polaris~Ab, with an orbital period near 30 years (Roemer, 1965; Kamper,
1996; Usenko et al. 2005b). The main-sequence companion is
0$\lefteqn{.}^{''}176$ from the Cepheid primary according to recent
obervations by Evans et al. (2007). Two other optical companions,
Polaris~C and D (13 and 12$^m$, angular separations from Polaris~A of
43$^{''}$ and 83$^{''}$, respectively, are also potential members of the
multiple system (Fernie, 1966), although recent observations refute that
suggestion (Evans et al., 2007). It would therefore be interesting to
establish the atmospheric parameters and chemical composition of component
B for comparison with those for the Cepheid Polaris itself --- it is known
that Cepheids progenitors are main-sequence $B$-stars, more massive and
rapidly evolved stars than their $F$--type companions. It is particularly
importnat to examine the ``key elements'' for yellow supergiant evolution
--- the CNO elements, sodium, and magnesium. The most detailed analysis of
the chemical composition for Polaris is that in our earlier study (Usenko
et al., 2005a). It is also an important exercise to establish the distance
and mass of Polaris~B using its spectroscopically-established
$T_{\rmn{eff}}$ and $\log g$ values, for comparison with the masses of the
close companions: Polaris~Ab and B.

\section{Observations, atmospheric parameters, radial and projected
rotational velocities}

High-resolution CCD spectrum of Polaris~B within the
wavelengh ranges 4555--6000\,\AA{} have been obtained on
June 2006 (HJD\,2453904.879) using 6\,m telescope BTA of the
Special Astrophysical Observatory of the Russian Academy of
Sciences with the high resolution echelle spectrograph NES
(Panchuk et al. 2002). In combination both with CCD 2048\,x\,2048
pixels and with an image slicer (Panchuk et al. 2003), NES
provides the resolutional power R$\approx$60,000 within the
wavelengh ranges 3500--6800\,\AA{}. Signal-to-noise ratio at the
continuum level in each of 25 spectral orders exceeds 70. A
Th--Ar lamp was used for the wavelength calibration.

We need to draw an attention for so called scattered light problem
in this case of Polaris~B. First Fernie (1966) was confronted with
difficulties in his photometrical observations when the scattered
light from the primary component ``..appreciably contaminated the
light of the eight magnitude secondary...under the relatively poor
seeing conditions that preveal in this climate''. Afterwards Kamper
(1996) noted that fibre-linked CCD observations are ``ideal'' since
the small circular entrance aperture makes their ``..easier to avoid
contamination from Polaris~A''.

We have a lot of calculations of the radial energy distribution along
stellar images of the 6\,m telescope for different values of seeing.
All calculations were made in approximation of a stellar images by the
Lorentzian profiles. For details of the procedure see the paper of
Diego (1985).
In particular, at seeing $\beta$\,=\,3$^{''}$ the weakening of the flux
of a star located at distance $d$\,=\,18$^{''}$ from the slit is equal to
X\,=\,6\,x\,10$^{-5}$ (10.5$^m$). We have just such a seeing in our
observing run.
It means that we can not register the scattering light of Polaris
since it is known that we may register radiation of double stars
if the difference of their magnitudes is not more than 2$^m$.

As known, Polaris~A is a slow rotating yellow supergiant with sharp and
narrow absorptional deep lines. Indeed, we do not see any spectral
features (narrow absorptions) of Polaris~A in the spectrum of Polaris~B
having very broad lines as a result of fast rotation. Therefore we can
verify that scattered light influence from the primary component is
insignificant.

Using the ECHELLE context of MIDAS modified by Yushkin \&
Klochkova (2005) for the case of observations with an image
slicer, we extracted the spectra from the CCD images, subtracted
bias, applied flat-field corrections, removed cosmic rays,
performed the wavelength calibration, and summed up individual
images. Further work with the spectra, including normalization to
the continuum level, wavelength calibration and equivalent width
measurements ($W_{\lambda}\leq160$ m\AA), was done using the
DECH20 package (Galazutdinov 1992).

\section{Main parameters}

\begin{figure}
\epsfig{figure=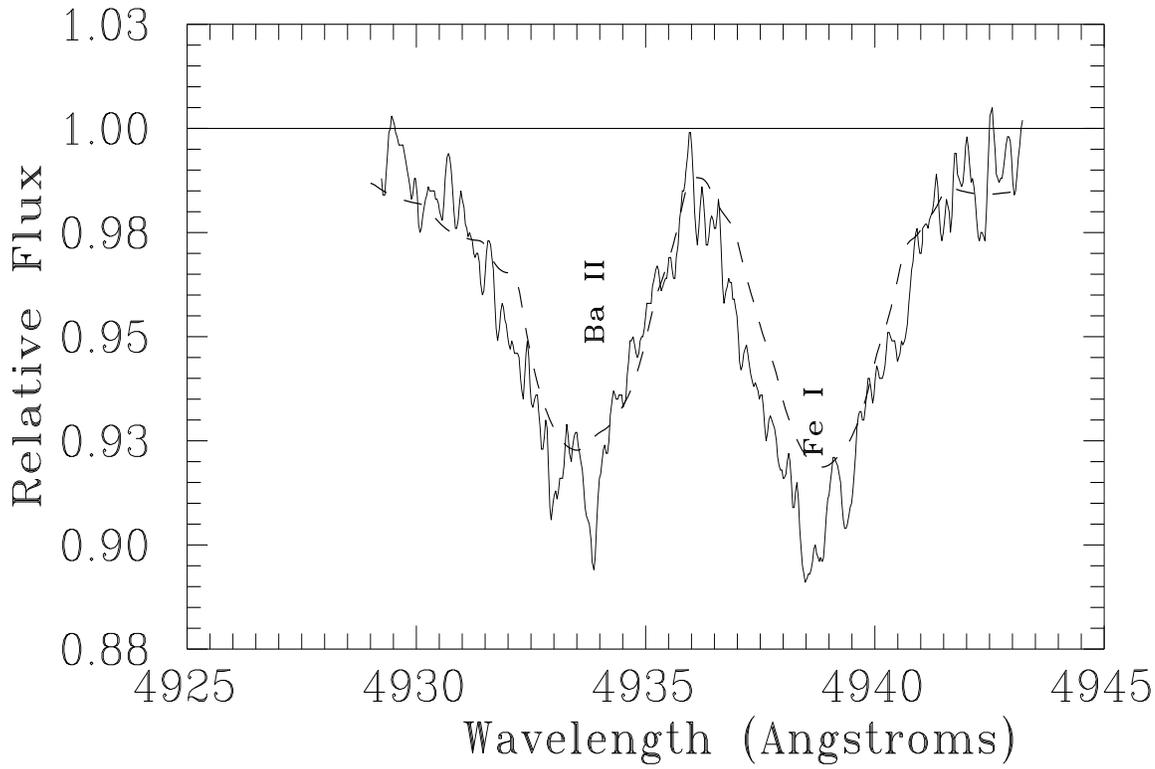,clip=,silent=,height=4in,width=6in}
\caption{Fragment of Polaris~B spectrum in range of $\lambda$$\lambda$
        4930--4943\,\AA\AA. Synthetic spectra (dashed line) calculated in case
        of
        of $T_{\rmn{eff}}$\,=\,6900\,K, $\log g$\,=\,4.3 and
    $v \sin i$\,=\,110\,km\,s$^{-1}$}
\label{Fig1}
\end{figure}

Before the atmospheric parameters determinations we made the visual
inspection of Polaris~B spectrum, because it has a relatively high
projected rotational velocity (see Fig.\,\ref{Fig1}.)

Radial velocity measurements were carried out using $H_{\beta}$ and 63
absorption lines of metals. The results are: $-16.6$\,km\,s$^{-1}$ and
$-18.9\pm$3.6\,km\,s$^{-1}$, respectively. These data are close within
the limits of errors to Kamper's (1996) result of
$-14.7\pm$1.2\,km\,s$^{-1}$.
To evaluate the line blends for Polaris~B, the spectrum synthesis
technique was applied. This was performed with the help of SYNSPEC code
(Hubeny et al. 1994). Projected rotational velocity was estimated by
fitting the synthesized spectrum to observed one.

\begin{figure}
\epsfig{figure=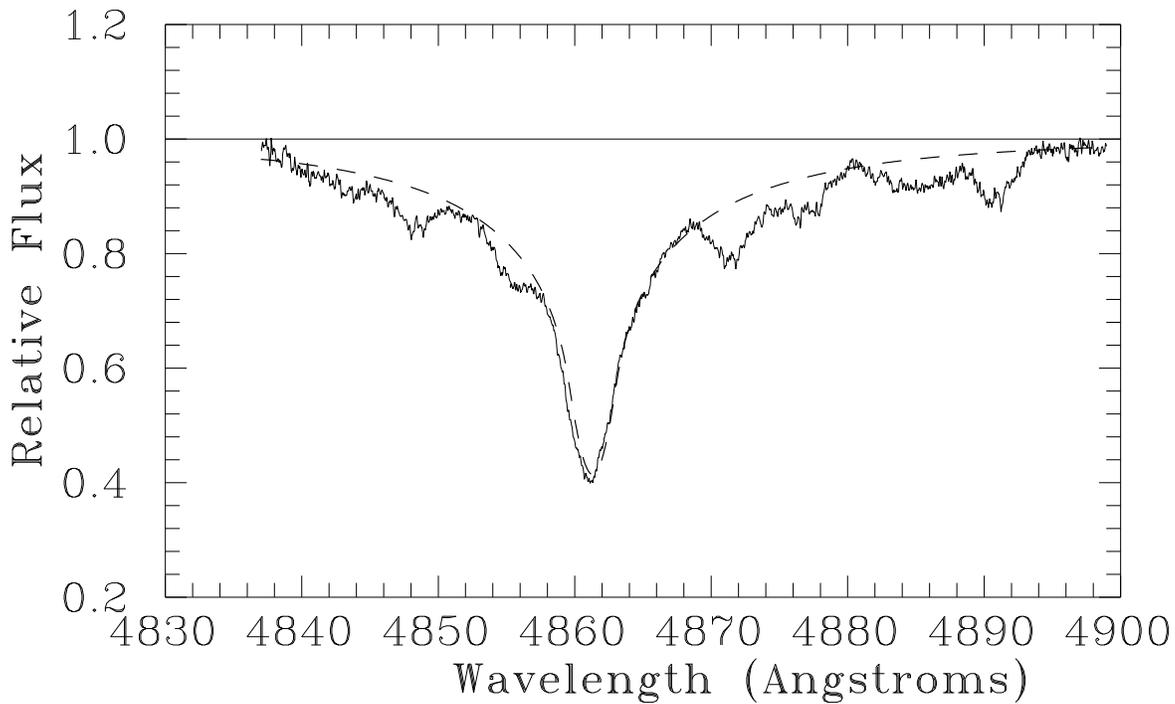}
\caption{The fit between observed and synthetic spectra (dashed line)
         for Polaris~B region near H$_{\beta}$ line in case of
     $T_{\rmn{eff}}$\,=\,6900\,K, $\log g$\,=\,4.3 and
     $v \sin i$\,=\,110\,km\,s$^{-1}$}
\label{Fig2}
\end{figure}

Before this procedure we have to set the preliminary estimates of
effective temperature and gravity values. Using $(B-V)$\,=\,0.42$^m$,
$(U-B)$\,=\,0.01$^m$ (Turner 2005) and $E_{B-V}$\,=\,0.034$^m$
(Usenko et al. 2005a) from (U-B),(B-V) -- T$_{\rmn{eff}}$, $\log g$
calibrations (Castelli 1991), the preliminary values were
$T_{\rmn{eff}}$\,=\,6900\,K and $\log g$\,=\,4.5. Starting with these
atmosphere parameters and to avoid the scattered light influence from
Polaris~A  we have use the synthetically generated
H$_{\beta}$ line profile and compared with the observed one. Finally we
have found the best fit in case of $v \sin i$\,=\,110\,km\,s$^{-1}$,
$T_{\rmn{eff}}$\,=\,6900 $\pm$ 50 K and $\log g$\,=\,4.3 $\pm$ 0.15
(see Fig.\,\ref{Fig2}).

The next step was to specify the surface gravity by adopting the
same iron abundance of the Fe{\sc i} and Fe{\sc ii} lines (with
a mean uncertainty of 0.15 dex). And, the microturbulent velocity
by assuming abundances of the Fe{\sc i} lines independent of the
equivalent width $W_{\lambda}$ (a mean error of 0.25\,km\,s$^{-1}$).
All the atmosphere models and chemical composition were calculated
using our version of the WIDTH9 code on the basis of the Kurucz (1992)
grid with the ``solar'' $\log {\rmn{gf}}$ values, adopted from Kovtyukh
\& Andrievsky (1999). As seen we had a happy choice for our second
shoot to detect the surface gravity, -- Fe{\sc i} and Fe{\sc ii}
abundances gave an equivalent results in case of $\log g$=4.3.

Finally, we have obtained for Polaris~B the following values of parameters:
T$_{\rmn{eff}}$\,=\,6900\,K; $\log g$\,=\,4.3;
$V_{\rmn{t}}$\,=\,2.5\,km\,s$^{-1}$;
$V_r$\,=\,$-16.6$\,km\,s$^{-1}$ (H$_{\beta}$) and $-18.9$\,km\,s$^{-1}$
(metallic lines); $v \sin i$\,=\,110 km\,s$^{-1}$.

\section{Chemical abundances}

Since Polaris~B spectrum displayed the absorptional lines spreading by a
high projected rotational velocity, we had to select unblended ones. In
Table~\ref{T1} the derived mean element abundances for Polaris~B are given
in comparison with ones of Polaris the Cepheid (Usenko et al. 2005a).

\begin{figure}
\epsfig{figure=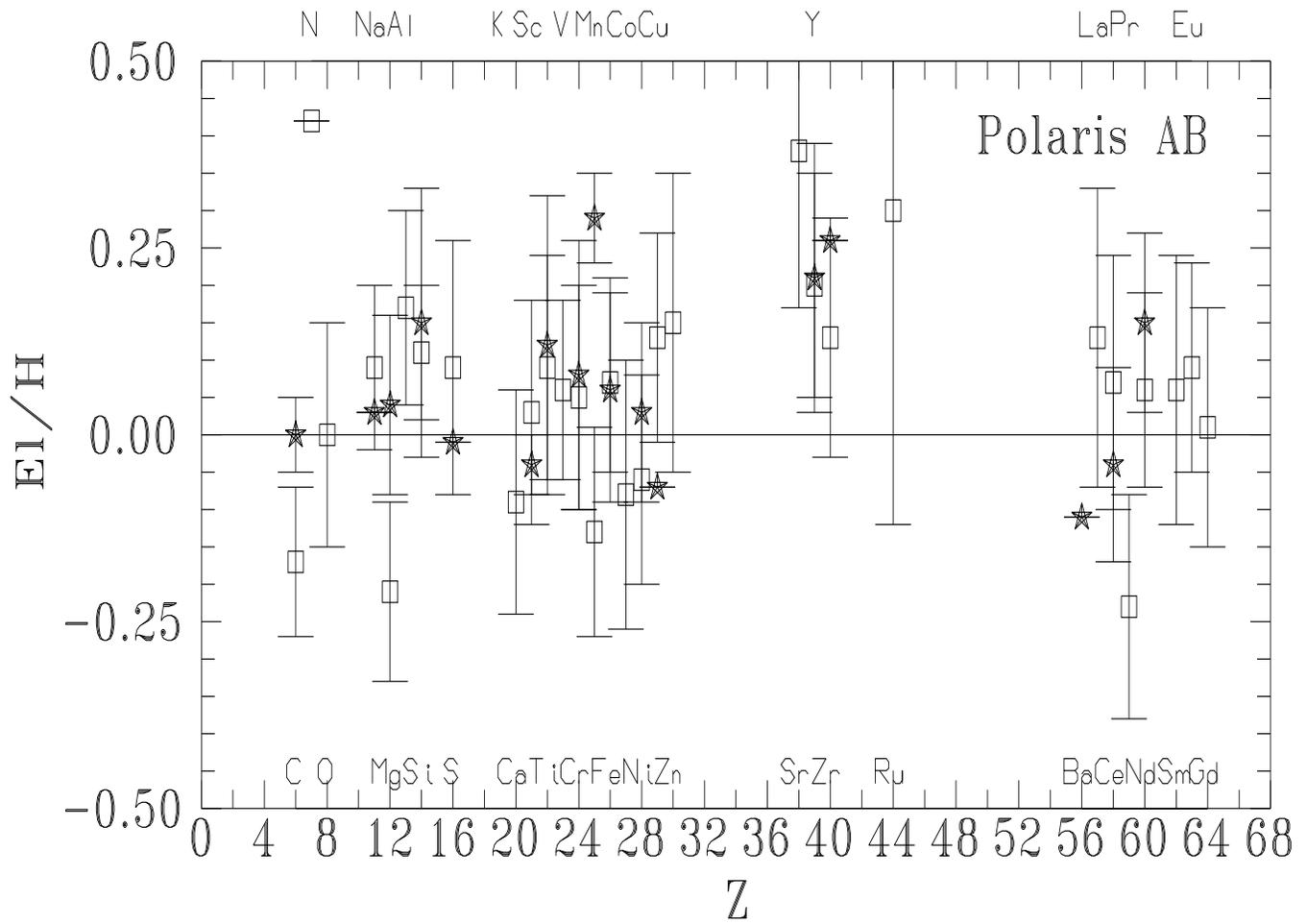,clip=,silent=,height=5in,width=7in}
\caption{Chemical abundances for Polaris~A (squares) and
          Polaris~B (filled stars)}
\label{Fig3}
\end{figure}

A comparison of chemical abundances of the both components displays
some interesting features. At about an equal (within the limits of
mean errors) iron, $\alpha-$, Fe-group (excepting manganese) and
s-process elements abundances for the both stars, Polaris~B
demonstrates solar carbon content, thereas Cepheid Polaris~A has
obvious deficit, conditioned by its possibly third or fifth Cepheids
instability strip crossing (Usenko et al. 2005a). The same cause is
noticeable for sodium and magnesium, -- its content for Polaris~B is
close to solar one too $(see Fig.\,\ref{Fig3})$.

\section{Radius, luminosity, distance and mass}

Since we can estimate a radius value for $F3\,{\sc v}$ spectral type
main-sequence near 1.38\,R$_{\sun}$ (Strai\v{z}ys 1982), therefore its
luminosity is 3.868\,L$_{\sun}$, absolute magnitude
M$_{\rmn{V}}$\,=\,+3.30$^m$. Using $A_{\rmn{V}}$\,=\,0.102$^m$ from Usenko
et al. (2005a) we can obtain the distance $D$\,=\,109.5\,pc. This result
has an ideal agreement with Kamper's (1996) one of 110\,pc, obtained by
astrometrical methods. According to Strai\v{z}ys (1982) the mass for such
star is equal to 1.31\,${\mathcal M}_{\sun}$, therefore $\log
(L/L_{\sun})$~$\approx$~4\,$\log (M/M_{\sun})$~ equation gives
1.4\,$M_{\sun}$. Using our $\log g$\,=\,4.3, and radius 1.38\,R$_{\sun}$
we can obtain a value of 1.39\,${\mathcal M}_{\sun}$. As seen, our mass is
satisfactory close to these two different sources of mass values.

\section{Conclusions}

We can summarize the results of our investigation as follows.
\begin{enumerate}
\item Radial velocity values for Polaris~B are close to
$\gamma$\,=$-15.9\pm$0.06\,km\,s$^{-1}$ of Polaris~A system.
\item High projected rotational velocity $v \sin i$\,=\,110\,km\,s$^{-1}$
is an evidence that system is young and Polaris~B is likely to be single,
since most binaries of $A-F$ types have slow rotation, the angular
momentum being tied up in orbital motion. Moreover, the rapid rotation's
observation could be mean that we see the star nearly equator-on.
\item Atmosphere parameters, obtained for Polaris~B are typical for
$F3{\sc v}$ star.
\item The majority of Polaris~B chemical elements shows abundances, equal
to Polaris~A and close to solar one. But carbon and sodium in Polaris~B
close to solar content, therefore Polaris~A demonstrates a typical for
the first dredge-up yellow supergiants deficit of C and Mg and
overabundance of Na (Usenko et al. 2005a). Therefore we are eye-witnesses
of evolutional history of two stars with different masses in the same
stellar system.
\item Absolute magnitude +3.30$^m$ is equal to one from Fernie (1966).
Spectroscopically determined effective temperature 6900\,K combinating
with radius of 1.38\,R$_{\sun}$ give the distance near 109.5\,pc  that
agrees with Kamper's (1996) 110\,pc one's. This result is quite
unexpected, because Turner (2005) denoted $101\pm$3\,pc to this object and
Polaris system as a whole. Whereas HIPPARCOS parallax (ESA 1997) and
optical interferometry (Nordgren et al. 1999) results give 132$\pm$9\,pc to
the Polaris~A.
\item The obtained mass of Polaris~B near 1.39\,${\mathcal M}_{\sun}$ has been founded
as unexpected close to one of Polaris~Ab spectroscopic companion, --
$1.38\pm0.61 M_{\sun}$ (Evans et al. 2007). The last one is, probably,
a main-sequence star of earlier than $F4{\sc v}$ spectral type (Evans et
al. 2002).
\end{enumerate}

\section{Acknowledgments}
Authors are grateful to D.\,G.~Turner for discussions and recommendations.
We are much indebted to V.\,E.~Panchuk and M.\,V.~Yushkin for their help at
observing run with 6\,m telescope.

\begin{table}
\caption[]{Average elemental abundances for Polaris~B in comparison to
Polaris~A (Usenko et al. 2005a)}
\label{T1}
\Large
\begin{tabular}{lccrcc}
\hline\noalign{\smallskip}
Element & \multicolumn{3}{c}{Polaris~B} & \multicolumn{2}{c}{Polaris~A}\\
\cline{2-4}
\cline{5-6}
Element & [El/H] & $\sigma$ & NL & [El/H] & $\sigma$\\
\noalign{\smallskip}\hline\noalign{\smallskip}
C {\sc i}   &$-0.00$& 0.05 &  4 &$-0.17$& 0.10 \\
Na {\sc i}  & +0.03 &  --  &  1 & +0.09 & 0.11 \\
Mg {\sc i}  & +0.04 & 0.12 &  2 &$-0.21$& 0.12 \\
Si {\sc i}  & +0.15 & 0.18 &  2 & +0.10 & 0.09 \\
S  {\sc i}  &$-0.01$&  --  &  1 & +0.09 & 0.17 \\
Sc {\sc ii} &$-0.04$& 0.04 &  2 &$-0.02$& 0.11 \\
Ti {\sc i}  & +0.13 & 0.20 &  3 & +0.09 & 0.17 \\
Ti {\sc ii} &$-0.01$& 0.21 &  2 & +0.05 & 0.08 \\
Cr {\sc i}  & +0.11 & 0.20 &  6 & +0.05 & 0.20 \\
Cr {\sc ii} & +0.02 & 0.16 &  3 & +0.06 & 0.13 \\
Mn {\sc i}  & +0.29 & 0.06 &  2 &$-0.13$& 0.14 \\
Fe {\sc i}  & +0.05 & 0.14 & 24 & +0.07 & 0.12 \\
Fe {\sc ii} & +0.08 & 0.25 &  3 & +0.07 & 0.08 \\
Ni {\sc i}  & +0.03 & 0.12 & 11 &$-0.06$& 0.14 \\
Cu {\sc i}  &$-0.07$&  --  &  1 & +0.13 & 0.14 \\
Y {\sc ii}  & +0.21 & 0.18 &  2 & +0.17 & 0.15 \\
Zr {\sc ii} & +0.26 &  --  &  1 & +0.13 & 0.16 \\
Ba {\sc ii} &$-0.11$&  --  &  1 &  --   &  --  \\
Ce {\sc ii} &$-0.04$& 0.13 &  2 & +0.07 & 0.17 \\
Nd {\sc ii} & +0.15 & 0.12 &  3 & +0.06 & 0.13 \\
\noalign{\smallskip}\hline
\end{tabular}
\begin{list}{}{}
\item[NL] -- number of lines
\end{list}
\end{table}

\clearpage
\newpage

\label{lastpage}

\end{document}